\documentclass[unsortedaddress, pre, 11pt, showpacs, amsmath, amssymb, superscriptaddress, flushbottom]{revtex4}
\usepackage{subfigure}
\usepackage{tikz}
\usetikzlibrary{patterns}
\usepackage{bm}
\usepackage{amsmath}
\usepackage{amssymb}
\usepackage{tabularx}
\usepackage{bbm}
\usepackage[utf8]{inputenc}
\usepackage[T1]{fontenc}
\usepackage{graphicx,color}
\usepackage[english]{babel}
\usepackage{times}
\usepackage{ae}
\usepackage{dcolumn}

\newcommand{\equref}[1]{Eq.~(\ref{#1})}

\newcommand{\Equref}[1]{Equation~(\ref{#1})}
\newcommand{\figref}[1]{Fig.~\ref{#1}}

\newcommand{\Reff}{\ensuremath{R_{\text{eff}}}}
\newcommand{\br}{\ensuremath{\bm{r}}}
\newcommand{\Ucm}{\ensuremath{U_{\text{cm}}}}
\newcommand{\Ycm}{\ensuremath{Y_{\text{cm}}}}
\newcommand{\PhiR}{\ensuremath{\Phi_{\text{R}}}}
\newcommand{\PhiT}{\ensuremath{\Phi_{\text{T}}}}
\newcommand{\sigmaLV}{\ensuremath{\sigma_{\text{LV}}}}
\newcommand{\thetaeq}{\ensuremath{\theta_{\text{eq}}}}
\newcommand{\thetaY}{\ensuremath{\theta_{\text{Y}}}}
\newcommand{\rhoL}{\ensuremath{\rho_{\text{L}}}}
\newcommand{\rhoV}{\ensuremath{\rho_{\text{V}}}}
\newcommand{\delete}[1]{}
\begin{document}

%-----------------------------------------------------------------------
\title{{\bf Dynamics of cylindrical droplets on flat substrate: Lattice Boltzmann modeling versus simple analytic models}}
\author{Nasrollah Moradi}
\email{nasrollah.moradi@rub.de}
\affiliation{\textit{ICAMS}, Ruhr-Universit\"at Bochum, Stiepeler Strasse 129, 44801 Bochum, Germany}
\author{Fathollah Varnik}
\affiliation{\textit{ICAMS}, Ruhr-Universit\"at Bochum, Stiepeler Strasse 129, 44801 Bochum, Germany}
\affiliation{Max-Planck Institut f\"ur Eisenforschung, Max-Planck Str.~1, 40237 D\"usseldorf, Germany}
\author{Ingo Steinbach}
\affiliation{\textit{ICAMS}, Ruhr-Universit\"at Bochum, Stiepeler Strasse 129, 44801 Bochum, Germany}

\date{\today}

%-----------------------------------------------------------------------

\begin{abstract}
\par

The steady state motion of cylindrical droplets under the action of external body force is investigated both theoretically and via lattice Boltzmann simulation. As long as the shape-invariance of droplet is maintained, the droplet's center-of-mass velocity linearly scales with both the force density and the square of droplet radius. However, a non-linear behavior appears as the droplet deformation becomes significant. This deformation is associated with the drop elongation occurring at sufficiently high external forcing. Yet, independent of either the force density  or the droplet size, the center-of-mass velocity is found to be linear in terms of the inverse of dynamic viscosity.  In addition, it is shown that the energy is mainly dissipated in a region near the substrate particularly close to the three phase 
 contact line. The total viscous dissipation is found to be proportional to both the square of force density and the inverse of dynamic viscosity. Moreover, the  dependence of the center-of-mass velocity on the equilibrium contact angle is investigated. A simple analytic model is provided reproducing the observed behavior.

\end{abstract}\maketitle

%----------------------------------------------------------------------

\section{Introduction}\label{action}
Individual droplets play a key role in many biological systems \cite{Wolgemuth}. Droplet behavior  is  also crucial for numerous industrial applications such as in automobile manufacturing and drug production as well as glass industry. Consequently,  understanding the underling physics behind droplet behavior and finding novel applications are currently an active field of research \cite{Quere, QuereAnnu}. Recently, study of microdroplets has received lots of attentions  both experimentally and by numerical modeling. For example,  droplet spreading  on chemically and topographically patterned substrates, droplet evaporation, and wetting properties of superhydrophobic surfaces   have been extensively studied in the literature \cite{ Varnik2, Markus, Lipowsky, Seemann, Lenz, Dorrer, Reyssat1, Ajdari}. Particularly, controlling droplet motion is essential for many industrial purposes ranging from microfluidic devices to fuel cells and inkjet printing \cite{Reyssat}. 

The equilibrium contact angle  of a droplet, $\thetaeq$, placed  on a perfectly flat and homogeneous solid substrate is given by the  Young equation, $\cos\thetaeq=(\sigma_{SV}-\sigma_{SL})/\sigma_{LV}$, where $\sigma_{LV}$, $\sigma_{SL}$, and $\sigma_{SV}$ are the surface tensions of liquid-vapor, solid-liquid and solid-vapor, respectively  \cite{Young}. However, in the case of  moving drops, the advancing contact angle is often found to be larger than the receding \cite{deGennes}. This deference can be considered as a measure of droplet deformation and it may appear  in characterizing of droplet velocity   \cite{Kim}. A droplet may move due to a wettability or temperature gradient \cite{Brochard, Varnik1, Thiele}. Recently, we were successful to report a spontaneous droplet motion on a substrate topographically patterned with a step-wise gradient of pillars \cite{Nasrollah}.  Obviously, a droplet may also move under the action of a body force, \textit{e.g.},  a falling  drop on an inclined surface  under the gravitational forcing. Depending on the material parameters of the considered system such as $\eta$, dynamic viscosity, and $\thetaeq$ as well as  superhydrophobicity of the substrate, droplets perform   a sliding, rolling, or tank treading motion or a combination thereof \cite{Hodges, Aussillous, Mahadevan}. Associated to a very high external forcing (\textit{i.e.} sufficiently large velocity), drops may highly be elongated (pearling) and, further up, they  exhibit a cuspid tail that emits smaller drops \cite{Podgorski}. Introducing slippage at solid boundary is another issue that helps to characterize droplet motion \cite{Muller}.  

Here, we concentrate on the steady state motion of cylindrical drops. However, despite the apparent simplicity of the problem, several issues, such as dependence of center-of-mass velocity, $\Ucm$, and the dissipation loss on the material parameters and external forcing as fully as the role of droplet deformation are still not well understood \cite{Muller}. The steady state is reached  due to the balance between the rate at  which energy is imparted onto the droplet and the rate of energy dissipation. In general, there are different possible mechanisms for energy dissipation within a moving droplet: the viscous dissipation due to the velocity gradients, dissipation at the vicinity of the three phase contact line, and  the dissipation in the  precursor film which may form around the droplet in contact with a solid \cite{Quere, deGennes}. Since, the numerical model used in the present studies does not take account of precursor film, we will focus on the effects related to dissipation only. This includes both the bulk of the drop as well as the vicinity of the contact line. Interestingly, as long as external force is sufficiently weak  or --equivalently-- the droplet volume is sufficiently low so that the drop approximately maintains its equilibrium shape during motion, the dependence of the center-of-mass velocity on external force and on the droplet volume can be easily worked out via simple rescaling of the relevant parameters. In particular, we find that the steady state drop velocity is directly proportional to $g \Reff^2 / \eta$, where $g$ is the external body force (equivalent of the gravitational acceleration), $\Reff$ the effective drop radius and $\eta$ the shear viscosity. Deviation from this simple behavior is observed in the case of strong droplet deformation. However, since dynamic viscosity does not affect the droplet shape, the drop velocity remains proportional to $1/\eta$ even in the strongly deformed limit. Using numerical simulations, we also calculate the local energy dissipation inside the droplet. It is observed that the main dissipation takes place within a volume below the drop's center-of-mass. Based on this observation, we propose a simple model which successfully captures the dependence of drop velocity on equilibrium contact angle.

\section{Numerical Model}
Because of complicated nature of fluid flows, tractable analytical approaches are  often limited to simplified systems. In addition,  experimental studies are available only for a restricted range of parameters. In this context, computer simulations can help to bridge the gap between analytical approaches and experiments. In the past two decades, the lattice Boltzmann (LB) method \cite{McNamara1988,Higuera1989a,Higuera1989,Benzi1992,Qian1992,Rothman,Succi2001,Wolf-Gladrow2000} has  proved itself as a powerful Navier-Stokes solver for simulating a wide range of complex fluidic systems.

We employ a free-energy-based two-phase lattice Boltzmann (LB) model to solve the discrete Boltzmann equation (DBE) for the van der Walls fluid with the BGK approximation. A detailed description of the model can be found in references \cite{Lee1,Lee2}. For the sake of completeness, however, a brief overview of the model is provided here.  The DBE with external force $\textbf{F}$ can be written as
\begin{equation}
\frac{\partial f_{\alpha}}{\partial t}+\mathbf{e}_{\alpha} \cdot \mathbf{\nabla}f_{\alpha}=  -\frac{f_{\alpha}-f_{\alpha}^{eq}}{\lambda}+ \frac{(\mathbf{e_{\alpha}}-\mathbf{u}) \cdot \mathbf{F}}{\rho c_{s}^{2}} f_{\alpha}^{eq}.
\label{eq1}
\end{equation}
In the above, $f_{\alpha}$, $ \mathbf{e}_{\alpha}$ and $\mathbf{u}$ are particle distribution function,   the microscopic particle velocity and  the macroscopic velocity, respectively. The parameter $\rho$ stands for the fluid density, $\lambda$ is the relaxation time and $c_{s}$ denotes the sound speed. The non dimensional relaxation time $\tau=\lambda/ \delta t$ is related to kinematic viscosity by $\nu=\tau c_{s}^{2}\delta t$. The equilibrium distribution function, $f_{\alpha}^{eq}$, is given by 
\begin{equation}
f_{\alpha}^{eq}=w_{\alpha}\rho \left[ 1+\frac{\mathbf{e_{\alpha} \cdot \mathbf{u}}}{c_{s}^{2}}+ \frac{(\mathbf{e_{\alpha} \cdot\mathbf{u})^{2} }}{2c_{s}^{4}}-\frac{\mathbf{u} \cdot \mathbf{u} }{2c_{s}^{2}} \right],
\label{eq2}
\end{equation}
where $w_{\alpha}$is a weighing factor. In order to eliminate the parasitic currents, 
the averaged external force experienced by each particle $\mathbf{F}$ is chosen in the  potential form
\begin{equation}
\mathbf{F}=\mathbf{\nabla}\rho c_{s}^{2}-\rho\mathbf{\nabla}(\mu_{0}-\kappa \mathbf{\nabla}^{2}\rho),
\label{eq3}
\end{equation}
where $\mu_{0}$ is the chemical potential and $\kappa$  the gradient parameter. The equilibrium properties of the present model can be obtained from a free-energy functional consisting of a volume and a surface part,
\begin{equation}
\Psi=\int_{V}\left(E_{0}(\rho)+\frac{\kappa}{2}|\mathbf{\nabla}|^{2}\right)dV-\int_{S}(\phi_{1}\rho_{s})dS,  
\label{eq4}
\end{equation}
where $V$ is the system volume and $S$ the surface area of the substrate. The bulk energy density, $E_{0}$, can be approximated by $E_{0}(\rho)=\beta(\rho-\rhoV)^{2}((\rho-\rhoL)^{2})$ in which $\beta$ is a constant and both  $\rhoL$  and $\rhoV$ are saturation densities in liquid and vapor phase, respectively.  The  gradient parameter and  the liquid-vapor surface tension   can be computed as $\kappa=\beta D^{2}(\rhoL-\rhoV)^{2}/8$  and        $ \sigma=(\rhoL-\rhoV)^{3}\sqrt{2\kappa\beta}/6$, respectively.  The interface thickness $D$, $\beta$, $\rhoL$,  and $\rhoV$ are input parameters. The  second integral in \equref{eq3} is the contribution of solid-liquid interfaces in the total free energy $ \Psi$. At equilibrium, there are two solutions that satisfy  $\phi_{1}=\pm\sqrt{2\kappa E_{0}(\rho)}$.  Minimizing the free energy functional  $\Psi$  leads to an equilibrium boundary condition for the spatial derivative of fluid density in the direction normal to the substrate $\partial \bot \rho= - \phi_{1}/\kappa$. The parameter $\phi_{1}$ is related to  $\thetaeq$ via
\begin{eqnarray}
\phi_{1}&=\frac{\sqrt{2\kappa\beta}}{2} (\rhoL-\rhoV)^{2}\text{sgn} \left(\frac{\pi}{2}-\thetaeq\right)\nonumber\\
& \times 
\left\{ \cos\left(\frac{\alpha}{3}\right)\left[1-\cos\left(\frac{\alpha}{3}\right)\right]\right\}^{1/2},
\label{eq5}
\end{eqnarray}
where $\alpha=\text{arccos(sin}\thetaeq)^{2}$. \\

\begin{figure*}[!ht]
\includegraphics[ width=5cm]{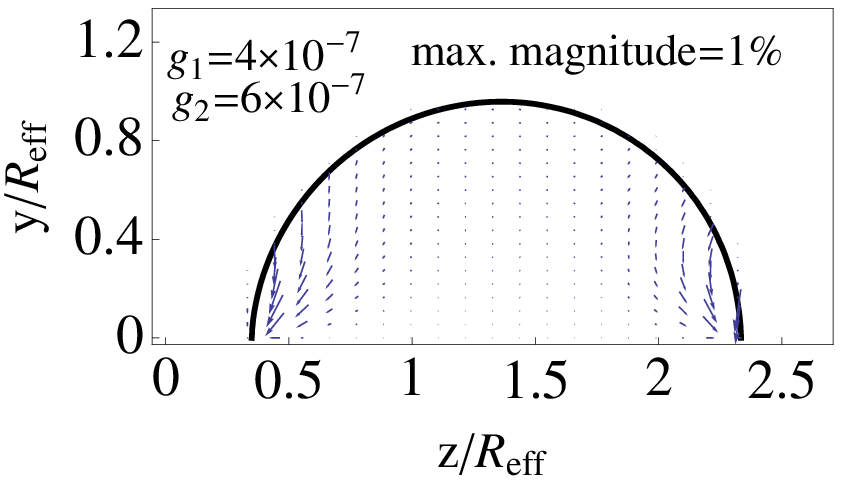}\hspace*{5mm}
\includegraphics[ width=5cm]{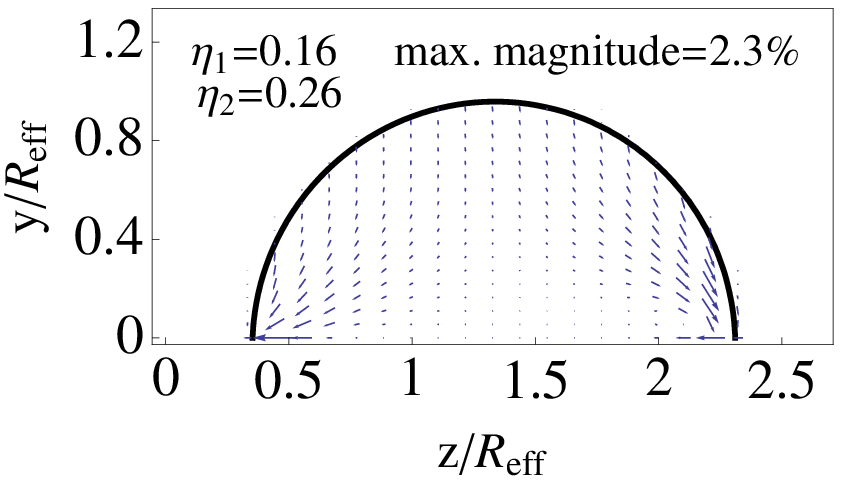}\hspace*{5mm}
\includegraphics[ width=5cm]{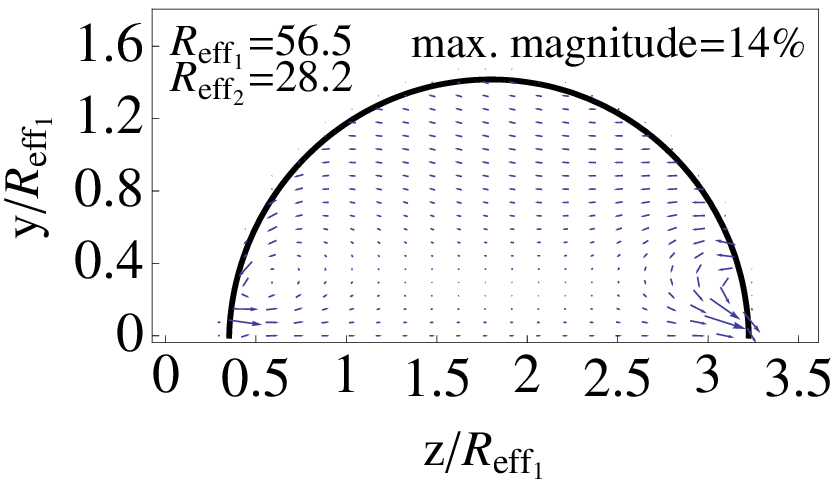}
\caption{(color on-line) Difference between rescaled velocity fields, $\hat{u}_2(\hat{y},\hat{z})-\hat{u}_1(\hat{y},\hat{z})$, for two different values of $g$ (left), $\eta$ (middle) and $\Reff$ (right) as indicated. Other control parameters of the simulation are as follows:  $\eta=0.16$ and $\Reff=19.7$ (left panel); $g=10^{-7}$ and $\Reff=19.7$ (middle panel) and finally $g=5\times10^{-8}$ and $\eta=0.16$ (right panel).  The difference between rescaled velocity fields in computed after a shift operation such that the center-of-mass of the droplets coincide with one another.}
\label{fig_comparing_velocity_profile}
\end{figure*}

The advantage of this model for the current study is both the possibility of achieving a high density ratio and, as it was already mentioned, the elimination  of parasitic currents at the liquid-vapor interface. It is important to note that the elimination of the spurious currents is an important step towards a reliable description of fluid dynamics inside a droplet. Simulating a two-phase system with a high density ratio, on the other hand, not only is more realistic but also allows to significantly reduce the finite size effect related to the dissipation loss in the vapor phase.

In our LB simulations, the bounce-back rule is imposed at solid boundaries.  For the open boundaries (in the $x$ and $z$-directions), the periodic boundary condition is applied. A body force, $\rho g$,  is applied to the liquid phase along the $z$-direction. The body force, however,  monotonously decreases through the interface and it vanishes in the gas phase. This accounts for the fact that the gas remains inert (static equilibrium) in the limit of zero droplet size. All the quantities in this paper are given in dimensionless LB units. The parameter $\beta$, the interface thickness $D$ and the saturation densities are fixed to $0.01$, $5$, $1$, and $0.01$, respectively. This choice of the parameters leads to a surface free energy of $\sigma \simeq 0.004$. Note that, in order to focus on situations, which can be easily controlled in real experiments, we do not change surface tension or liquid density in our simulations. Depending to the case of interest, the parameters $\tau$, $\thetaeq$, $\Reff$, and $g$ lie in the ranges $[0.02, 1.6]$, $[35^{\circ}, 140^{\circ}]$, $[22, 75]$ and  $[10^{-7}, 10^{-5}]$ in the order given. Typically, we use a simulation box of size $L_{x}\times L_{y} \times L_{z}$ $=$ $2\times120\times120$ lattice nodes. However, for large droplets, we increase the size of the simulation box (in the $y$ and $z$-directions) ensuring that there are no finite size effects. The volume of droplet is given by $V=SL_{x}$ where $S$ is the surface of droplet's cross-section normal to the $x$-direction. For the cylindrical geometry considered in this study, we define the droplet's effective radius as $\Reff=(S/\pi)^{1/2}$.

\section{A simple scaling relation}

Here, we investigate the effect of external forcing on the steady state motion of cylindrical drops on a flat surface. In addition, the influence of system parameters such as  droplet size $\Reff$, viscosity $\eta$ as well as equilibrium contact angle $\thetaeq$ on the steady state velocity of the droplet's center-of-mass is addressed.

By driving a droplet via an external body force, we mimic a real situation in which a droplet moves downward on an inclined surface due to the gravity. The external force does work on droplet with a rate equal to the total force applied on  the droplet multiplied by the droplet's center-of-mass velocity, $g \rho V \Ucm$. In the steady state, this energy is entirely transferred into dissipation. On the other hand, the total viscous dissipation is given by $\int_{V} S_{ij}\sigma_{ij} dV = \int_{V} \sigma_{ij}\sigma_{ij}/(2\eta) dV$, where the strain rate and stress tensors are given by $S_{ij} = (\partial u_{i}/\partial x_{j} + \partial u_{j} / \partial x_{i})/2$ and $\sigma_{ij}  = 2 \eta S_{ij}$, the later relation being valid for non-diagonal (shear) components of a Newtonian fluid (note that, due to the incompressibility of the liquid phase, the diagonal components of the strain rate and stress tensors are not relevant here). One thus obtains

\begin{equation}
g\rho V \Ucm = \frac{1}{2\eta} \int \sigma_{ij}\sigma_{ij} dV=2\eta \int S_{ij}S_{ij} dV.
\label{eq8}
\end{equation}
As long as the shape of the droplet does not change, it is reasonable to take $\Reff$ as a characteristic length. We also chose $\Ucm$ as a characteristic velocity and introduce dimensionless quantities such as $\hat{x}=x_{\alpha}/\Reff$  and  $\hat{u}_{\alpha}= u_{\alpha}/\Ucm$. Using these rescaled quantities, the strain rate tensor can also be written as $S_{ij}=\Ucm/\Reff (\partial \hat{u}_{i}/\partial \hat{x}_{j} + \partial \hat{u}_{j} / \partial \hat{x}_{i})/2=\Ucm/\Reff \hat{S}_{ij}$. Inserting this into \equref{eq8} yields 
\begin{equation}
g\rho V \Ucm = \frac{2 \eta \Ucm^2 V}{\Reff^2} \int  \hat{S}_{ij}\hat{S}_{ij} d\hat{V}.
\label{eq:balance}
\end{equation}
where we also made the volume element dimensionless ($dV=Vd\hat{V}$). The important step is now to assume that the {\it rescaled} velocity field within the droplet does not change upon a variation of the external force, drop radius or viscosity provided that the shape of the droplet remains constant. With this assumption, the integral in \equref{eq:balance} becomes a constant \textquoteleft shape factor\textquoteright\, and one obtains
\begin{equation}
\Ucm  \propto \frac{g \rho \Reff^2}{\eta}.
\label{eq:Ucm}
\end{equation}

It is noteworthy that, in the above model, the dependence of $\Ucm$ on $\Reff^2$ arises from the rescaling of the strain rate tensor $S^2_{ij}$ only. In particular, it remains valid regardless of the dimensionality of the space. Interestingly, when expressed in terms of droplet volume, $V$, the spatial dimension, $d$, does play a role. This is simply a consequence of the fact that  $\Reff \propto V^{1/d}$. In particular, $\Ucm  \propto V$ in 2D, while $\Ucm \propto V^{2/3}$ in 3D.

In order to test the above assumption of the scale invariance, we have performed a series of lattice Boltzmann simulations while varying $g$, $\eta$ and $\Reff$ in a range where droplet shape remains unchanged. The simulated velocity fields are then compared with one another by first rescaling the relevant velocity and length scales (see the text below \equref{eq8}) and then plotting the difference of the thus obtained velocity fields. 

Results of such an analysis are illustrated in \figref{fig_comparing_velocity_profile}. As seen from this figure, the rescaled velocity fields are very close to each other almost in the entire droplet with deviations in the vicinity of the three phase contact line. Noting that these deviations (being at most of the order of 10\%) are limited to a small fraction of the droplet's volume, the relative contribution of these deviations to the integral in the right hand side of \equref{eq:balance} becomes quite negligible in all the cases shown. Obviously, the assumption of a scale invariant velocity field is a good approximation to the actual flow behavior in the studied parameter range. \Equref{eq:Ucm} is thus expected to well describe our data as long as droplet shape is unaltered. 

\begin{figure}[t!]
\includegraphics[ width=6cm]{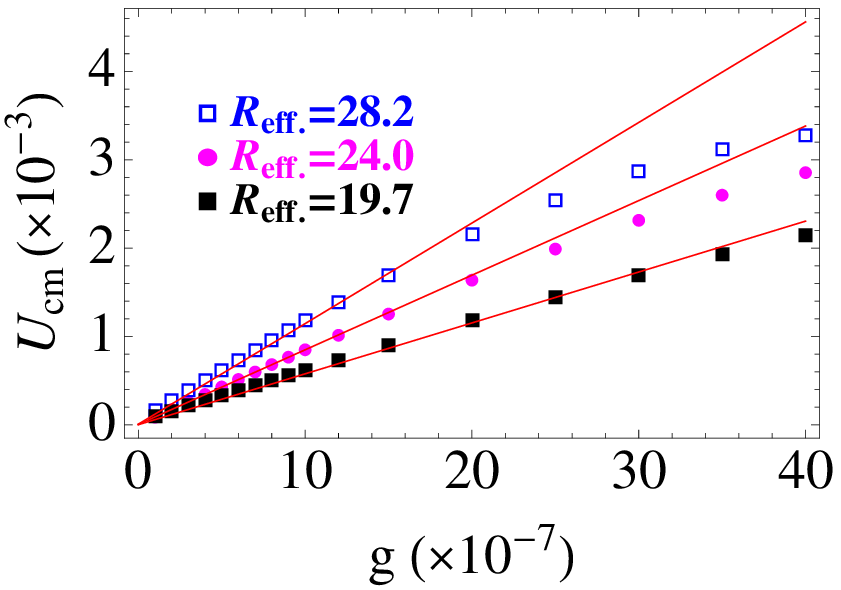}\hspace*{5mm}
\includegraphics[ width=6.4cm]{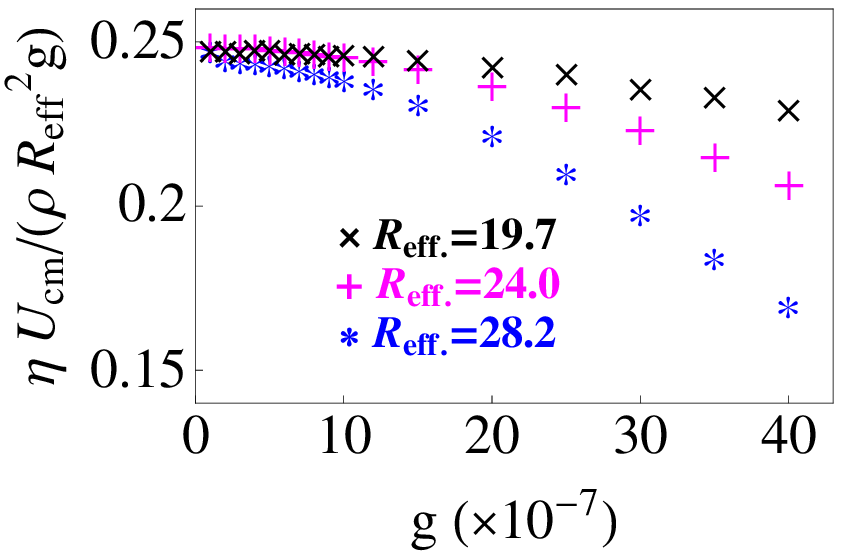}
\caption{(Color on-line) $\Ucm$  versus body force $g$ for different values of the effective droplet radius $\Reff$ as specified. A linear behavior is visible at sufficiently low $g$. The range of the validity of this linear regime is progressively restricted as droplet radius increases. In the right panel, $\eta \Ucm/(g\rho\Reff^2)$ is plotted versus $g$ for exactly the same data as in the left panel. In all these simulations, shear viscosity and equilibrium contact angle are set to $\eta=0.16$ and $\thetaY=90^{\circ}$.}
\label{fig_forcing_dependence}
\end{figure}

\begin{figure}[t!]
 \includegraphics[ width=6cm]{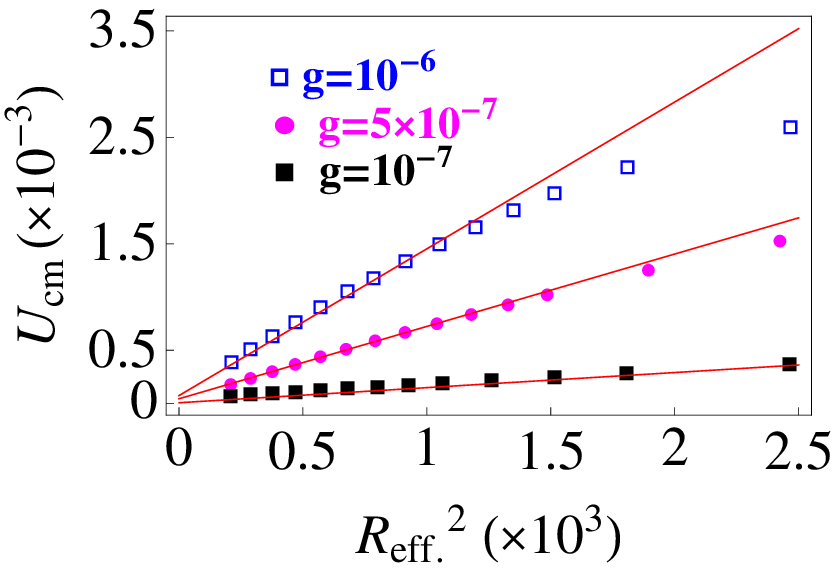}\hspace*{5mm}
 \includegraphics[ width=6.4cm]{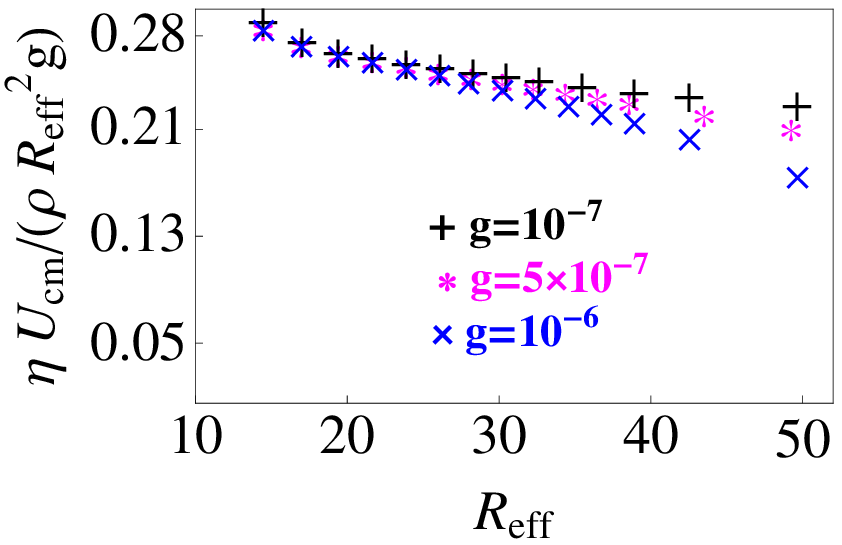}
\caption{(Color on-line)
$\Ucm$  versus $\Reff^2$ for different choices of the body force $g$ as indicated. Again, a linear behavior is visible at sufficiently low $\Reff$. The range of the validity of the linear behavior shrinks upon a raise of the body force. Following the same idea as in the  right panel of \figref{fig_forcing_dependence}, we plot in the right panel $\eta \Ucm/(g\rho\Reff^2)$ versus $\Reff^2$ for exactly the same data as in the left panel. In all these simulations, shear viscosity and equilibrium contact angle are set to $\eta=0.16$ and $\thetaY=90^{\circ}$.}
 \label{fig_size_dependence}
\end{figure}

The presence of a parameter range for the validity of \equref{eq:Ucm} is evidenced in \figref{fig_forcing_dependence}, where the center-of-mass velocity, $\Ucm$, is depicted versus force density, $g$, for different droplet sizes. As seen from this figure, the range of the validity of scaling relation \equref{eq:Ucm} extends to larger $g$ as droplet size decreases. Conversely, the larger the droplet, the earlier the onset of significant deviations. A similar trend is also observed in \figref{fig_size_dependence}, where droplet size is varied as control parameter for three different choices of $g$.

%-------------------------------------------------------------

\begin{figure*}[!ht]
(a)\includegraphics[width=4.6cm]{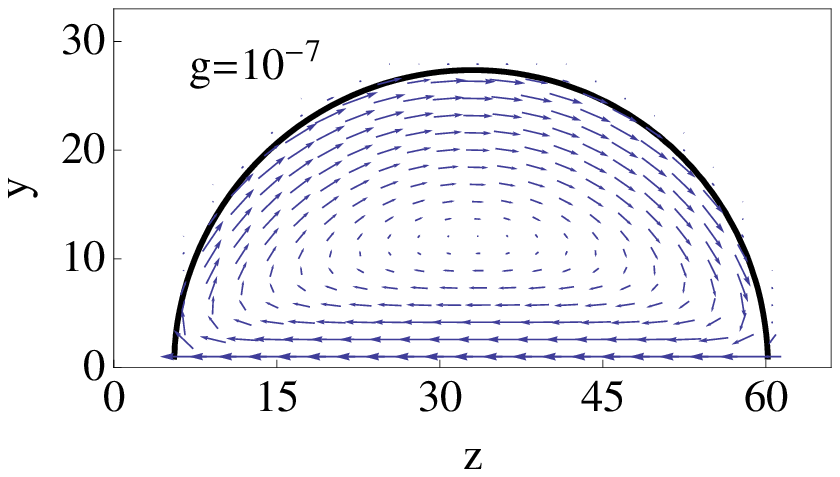}
(b)\includegraphics[width=4.7cm]{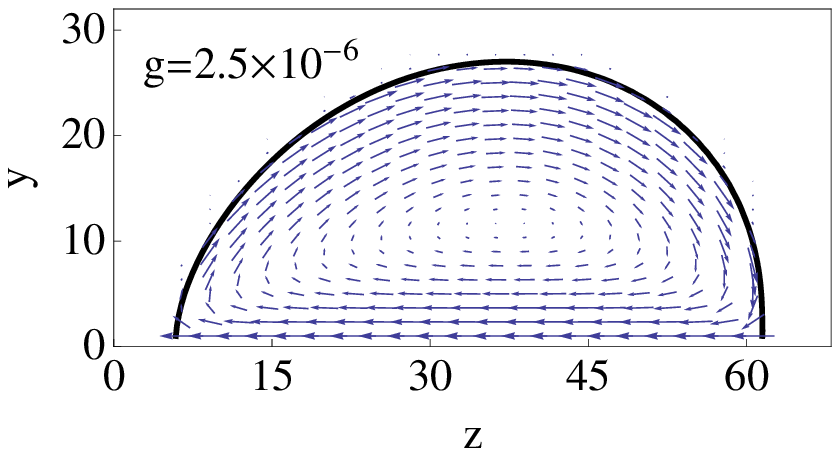}
(c)\includegraphics[width=4.8cm]{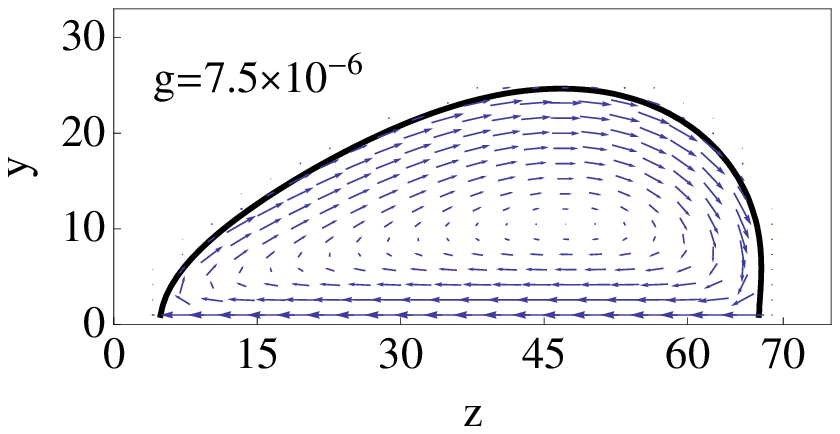}
\caption{Droplet shape and the corresponding momentum field  in the center-of-mass frame for three different values of the driving force $g$. As $g$ increases, the deformation becomes more pronounced. In the left panel, the deformation is negligible and the droplet's center-of-mass velocity obeys the simple relation \equref{eq:Ucm} for driving forces below the specified value. The middle panel marks the onset of deviations from \equref{eq:Ucm} and the left panel is well beyond the validity of this simple scaling relation. A rolling motion is clearly visible regardless of the deformation state  of droplet. In all the cases shown, the droplet's effective radius, dynamic viscosity and equilibrium contact angle are fixed to $\Reff=19.7$, $\eta=0.16$ and $\thetaY=90^{\circ}$, respectively. Recalling that $\sigmaLV=0.004$ and $\rhoL=1.0$, the Bond number from left to right reads $Bo=g\rho \Reff^2 / \sigmaLV \approx 1\times10^{-2},\;\; 0.25$ and $0.72$.}
\label{fig_drop_shape_rolling}
\end{figure*}

In the present study, the shape of droplet is determined by the competition between the surface force and the total body force. For a cylindrical drop of the cross sectional radius $\Reff$ and axial length $L_x$, this leads to $\sigmaLV L_x \le g \rho \Reff^2 L_x$ as a condition for a significant deformation. By introducing the Bond number, $Bo=\rho g \Reff^{2}/\sigmaLV$, one sees that strong deformation is expected for $Bo \ge 1$. Within prefactors of the order of unity, the same condition for drop deformation is also obtained in the case of a spherical droplet (to see this, replace $L_x$ by $\Reff$). It must be emphasized here, that this criterion is based on a scaling argument and the precise value of the Bond number for the transition from undeformed to a deformed state may be different from unity. What is essential here is the fact that a higher Bond number leads to a higher degree of deformation. In the case of our simulations, for example, slight but observable deformation occurs already for a Bond number as low as 0.25 (\figref{fig_drop_shape_rolling} b) with a significant increase in the deformation state as Bo increases from 0.25 to 0.72 (\figref{fig_drop_shape_rolling}c).

Droplet shapes and the corresponding momentum fields are shown in \figref{fig_drop_shape_rolling} for three typical values of $g$. As seen from the left panel of this figure, for a sufficiently weak body force (here $g=10^{-7}$), the deformation of the droplet is quite negligible but it becomes important upon an increase of $g$ (middle and right panels). 

Furthermore, \figref{fig_drop_shape_rolling} also shows the momentum field inside the droplet providing direct evidence for the existence of rolling motion in the center-of-mass frame of reference. Thus, an observer moving with the droplet's center-of-mass will confirm the presence of a well established rolling motion inside the droplet regardless of its deformation state. This rolling motion is associated to the tendency of droplet to minimize its total  dissipation loss  \cite{ Mahadevan, Aussillous}. Interestingly, similar rolling motion are also observed in molecular dynamics simulations of polymeric liquids \cite{Muller}.

\begin{figure}[t!]
\includegraphics[ width=6cm]{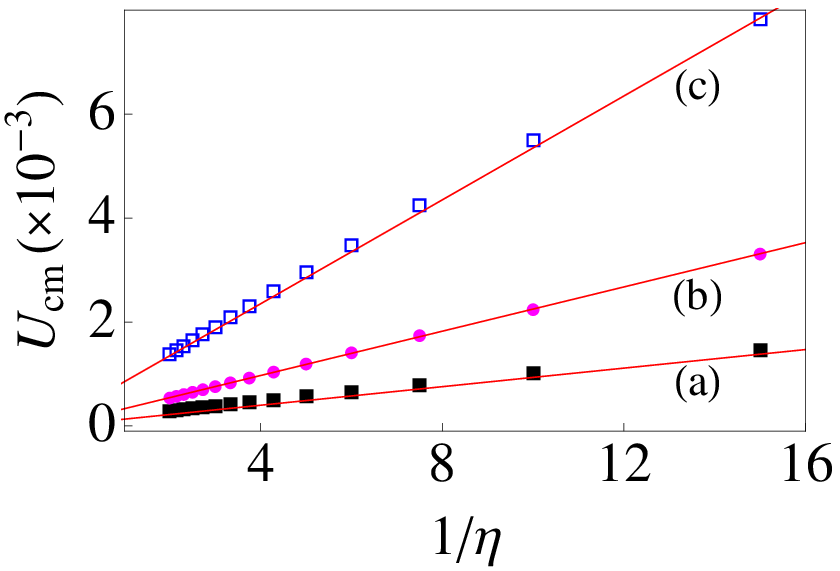}\hspace*{5mm}
 \includegraphics[ width=6.4cm]{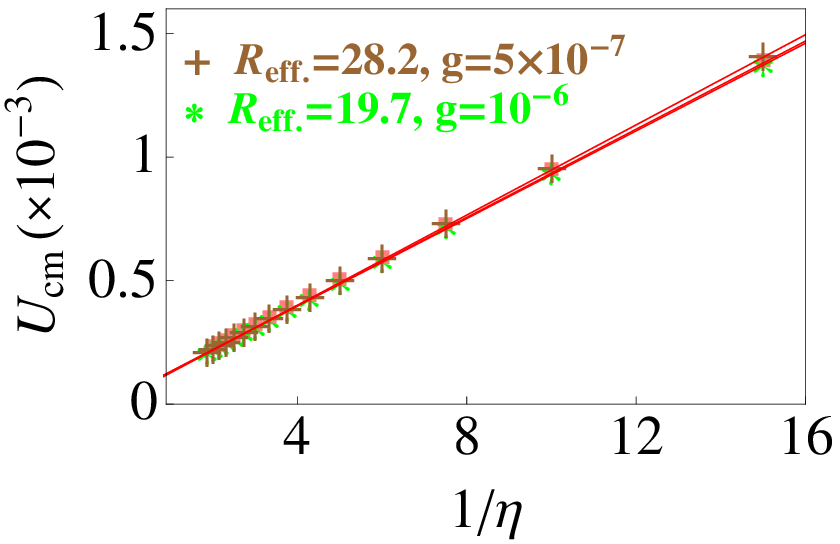}
\caption{ (Color on-line) $\Ucm$ versus  $1/\eta$ for  droplets with different degrees of deformation. The labeles (a)-(c) refer to deformation states shown in \figref{fig_drop_shape_rolling}, respectively (the velocity in the case (a) has been multiplied  by a factor of 10 for better visibility). In all the  cases shown, a perfect linear variation is seen in accordance with \equref{eq:Ucm}. In the left panel, we plot simulation results for two different choice of $\Reff$ and $g$  but keeping the product $\Reff^{2}g$ almost unchanged. In this case, the velocities of both droplets  fall onto a single line  which  also confirms the validity of \equref{eq:Ucm}.     $\thetaeq$ is fixed to $90^{\circ}$ in all cases.}
 \label{fig_viscosity_dependence}
\end{figure}

We close this section by addressing the effect of viscosity on $\Ucm$. For this purpose, we mention that a change in viscosity only affects the time scale of the entire simulation. In particular, a variation of viscosity has no influence on the shape of droplet. Consequently, we expect $\Ucm \propto 1/\eta$ regardless of the deformation state of droplet. This expectation is confirmed in \figref{fig_viscosity_dependence}, where $\Ucm$ versus $1/\eta$ is shown for droplets with different degrees of deformation.

%---------------------------------------------------------------------
\section{Local viscous dissipation}

In this section,  we provide a detailed analysis of the local dissipation rate, $\phi(\br)= \sigma^{2}(\br)/2\eta$  inside droplet. As will be shown hereafter, the insight gained via these investigations enables us to propose a simple model capable of accounting for the dependence of the total dissipation rate, $\Phi_T=\int \phi(\br)d^3\br$, on the contact angle, $\thetaY$. Equating this to the work done by the external force then yields a relation between the droplet's center-of-mass velocity and the equilibrium contact angle.

It is noteworthy that, unlike conventional Navier-Stokes solvers, the lattice Boltzmann method does not require ---although allows for--- the computation of velocity gradients to obtain the local stress tensor. Rather, it offers the unique possibility of obtaining the stress tensor \emph{locally} via the non-equilibrium part of the populations. In this regard, particular attention has been payed to a correct implementation of the stress computation \cite{Markus1}.

\begin{figure*}[!ht]
\unitlength=1mm
\begin{picture}(0,0)
\put(0,0){(a)}
\put(60,0){(b)}
\put(95,0){(c)}
\put(130,0){(d)}
\end{picture}
\includegraphics[ width=5.cm]{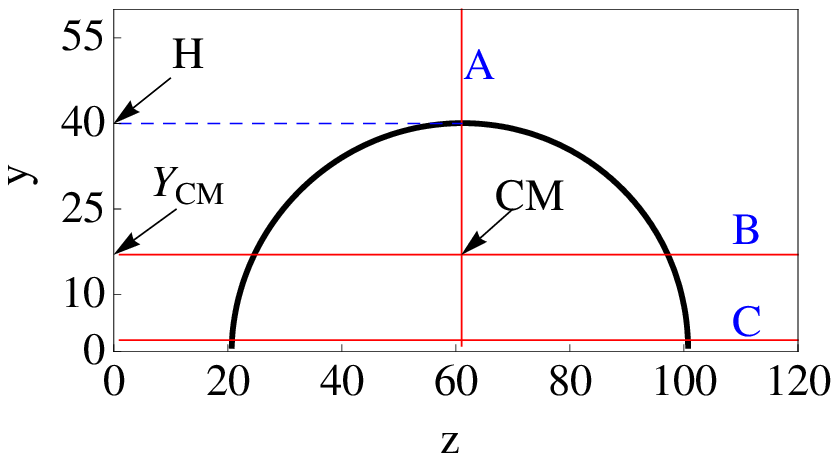}\hspace*{5mm}
\includegraphics[ width=3.cm]{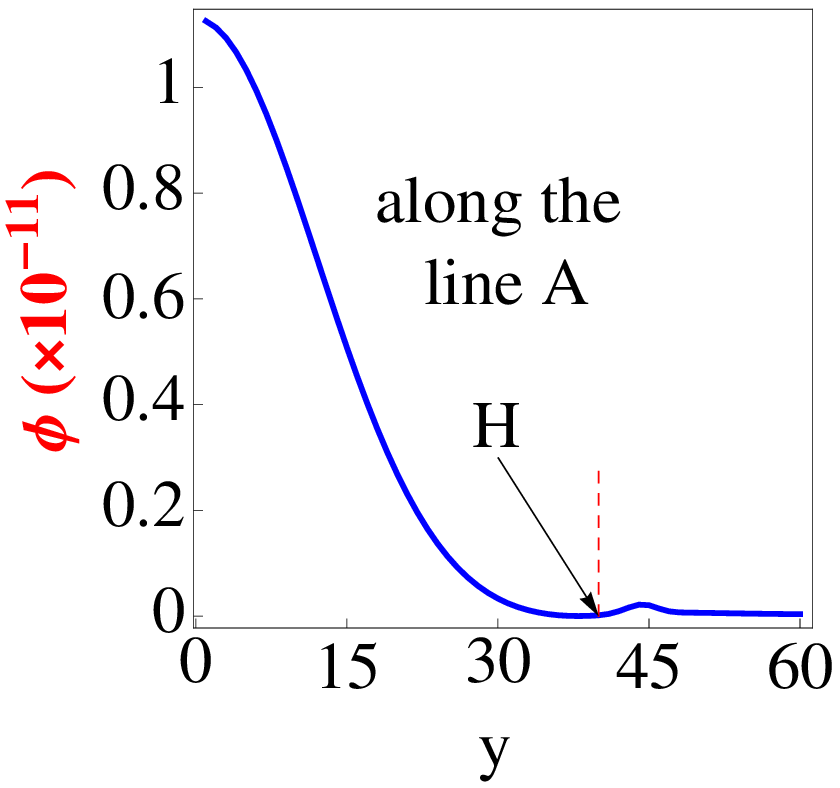}\hspace*{5mm}
\includegraphics[ width=3.cm]{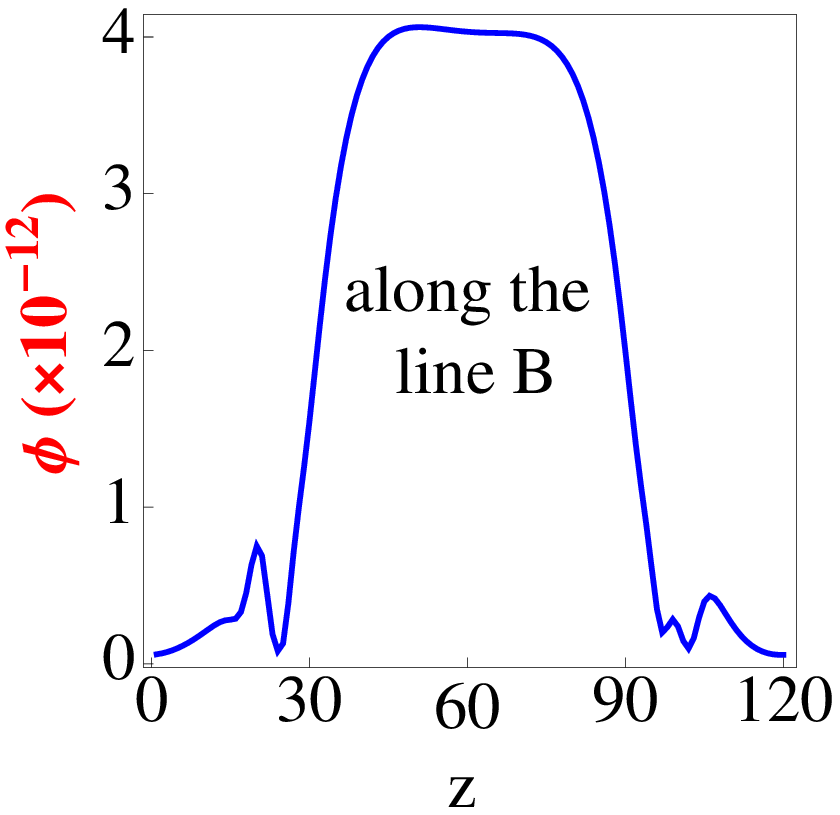}\hspace*{5mm}
\includegraphics[ width=3.cm]{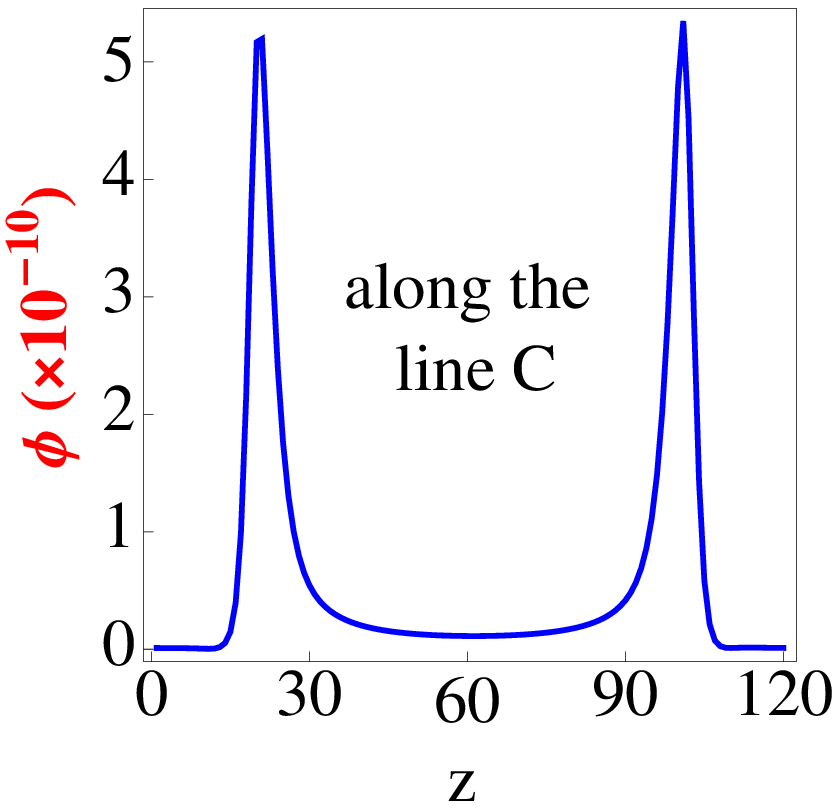}
\caption[]{(a) Illustration of the typical shape of a droplet and the lines along which viscous dissipation is determined (\textquoteleft CM\textquoteright\,  stands for the center-of-mass). The system parameters are $g=10^{-7}$, $\Reff=28.2$, $\eta=0.16$, and  $\thetaY=90^{\circ}$. (b-d) The variation of local viscous dissipation rate, $\phi=\sigma^2(\br)/(2\eta)$,  along lines A, B and C as indicated. Note that $\phi$ is  almost negligible in the gas phase. It has a large value close to the substrate, but rapidly  decreases far from the substrate. It  is also highly enlarged in the vicinity of triple contact line. }
\label{fig_viscous_dissipation_abc}
\end{figure*}

In order to figure out at which parts of droplet the energy is mainly dissipated, we compute local dissipation rate along the three lines labeled  by A, B, and C in the panel (a) of \figref{fig_viscous_dissipation_abc}. The variation of $\phi$ along these lines is depicted in the next panels of \figref{fig_viscous_dissipation_abc}. We first note that $\phi$ is negligible in the gas phase, which is often the case in real experiments due to the low vapor pressure. Furthermore ---as a comparison of the panels (a), (b) and (c) reveals--- the strongest dissipation occurs in the vicinity of the three phase contact line (panel (c)), which is roughly two orders of magnitude larger than the dissipation rate inside droplet (panel (b)). This behavior can be rationalized due to the fact that large velocity gradients occur near the substrate particularly in the vicinity of the three phase contact line \cite{Yeomans}. However, one must realize that bulk dissipation acts in a larger domain than the dissipation close to the contact line and thus may eventually dominate the overall dissipation rate if the droplet is sufficiently large.

\begin{figure}[!ht]
\includegraphics[ width=6cm]{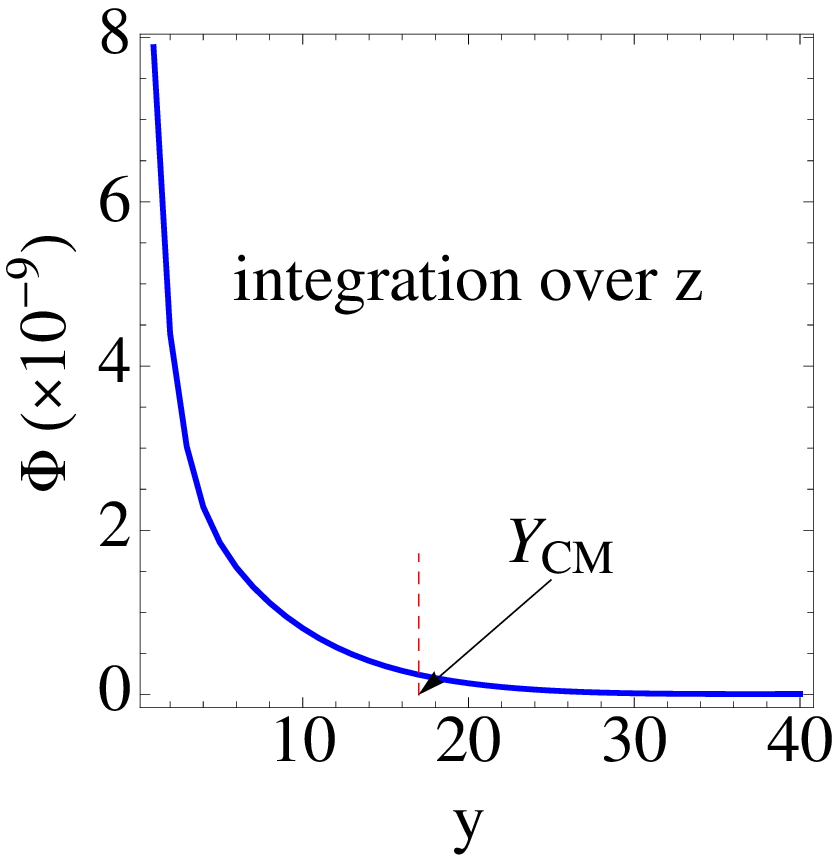}\hspace*{5mm}
\includegraphics[ width=6.4cm]{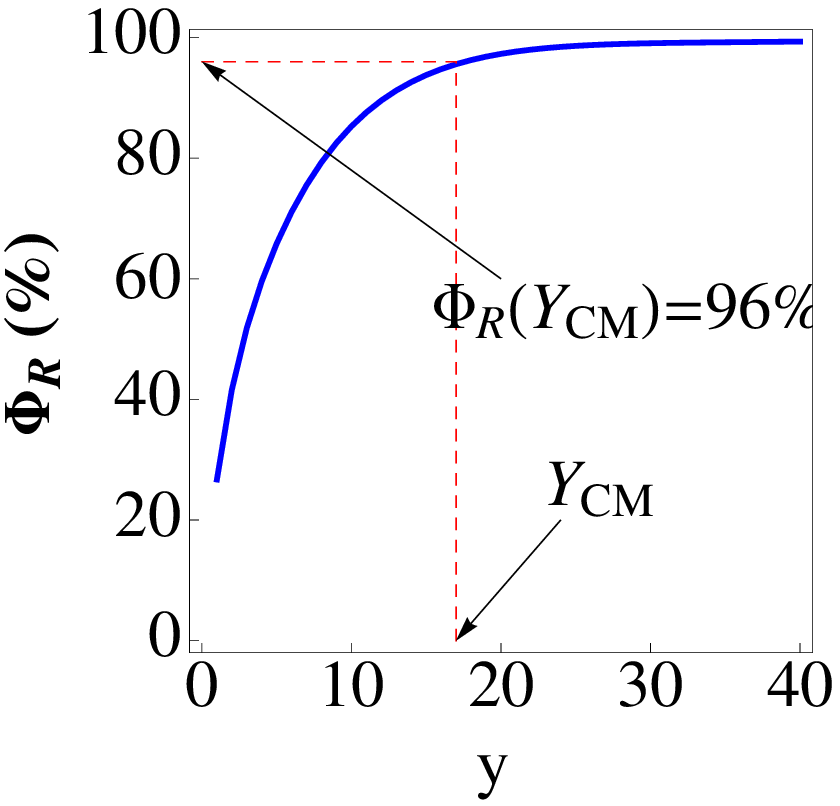}
\caption{ (Color on-line) variation of the local dissipation rate $\Phi$ and the relative dissipation rate  $\Phi_{R}$ as a function of $y$, corresponding to the droplet shown in \figref{fig_viscous_dissipation_abc}  in the left and  right, respectively. The main contribution to the total dissipation $\Phi_{T}$ occurs close to the substrate in a region given by $y< \Ycm$.}
\label{fig_viscous_dissipation_y}
\end{figure}

An interesting feature, relevant for our further analysis is the fact that viscous dissipation inside droplet is mainly localized to regions below the droplet's center-of-mass (panel (a) in \figref{fig_viscous_dissipation_abc}). This idea is further evidenced in \figref{fig_viscous_dissipation_y} (a), where we plot the viscous dissipation integrated along a horizontal line,  $\Phi(y)=\int_{0}^{Lz}\phi(y,z) dz$ as a function of vertical position $y$ (distance from the substrate). Indeed, as expected, viscous dissipation mainly occurs in a region specified by $y<\Ycm$. This finding is further underlined by showing in \figref{fig_viscous_dissipation_y} (b) the relative contribution, $\PhiR(y)$, to total dissipation within a region restricted between the substrate and a horizontal line at $y$,  $\Phi_{R}(y)=\int_{0}^{y}\Phi(y')dy' / \int_{0}^{Ly} \Phi(y')dy')$. It is visible  from \figref{fig_viscous_dissipation_y} (b) that $96\%$ of total dissipation occurs in a region specified by $y<\Ycm$.


Before working out an important consequence of this observation, we first check whether it remains valid upon a variation of shear viscosity and driving force. Inserting \equref{eq:Ucm} in the right hand side of \equref{eq:balance}, it is seen that the total dissipation rate is expected to obey 
\begin{equation}
\PhiT \propto  \frac{2 g^2 \rho^2\Reff^{2+d}}{\eta}.
\label{eq:PhiT}
\end{equation}

\begin{figure}[!ht]
\includegraphics[ width=6.3cm]{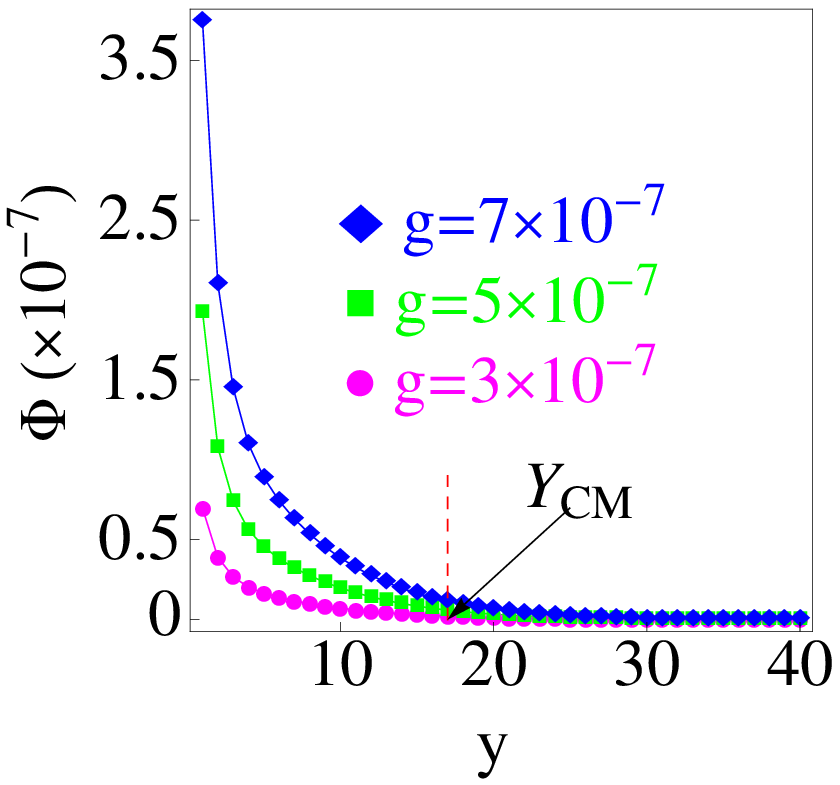}\hspace*{5mm}
\includegraphics[ width=6cm]{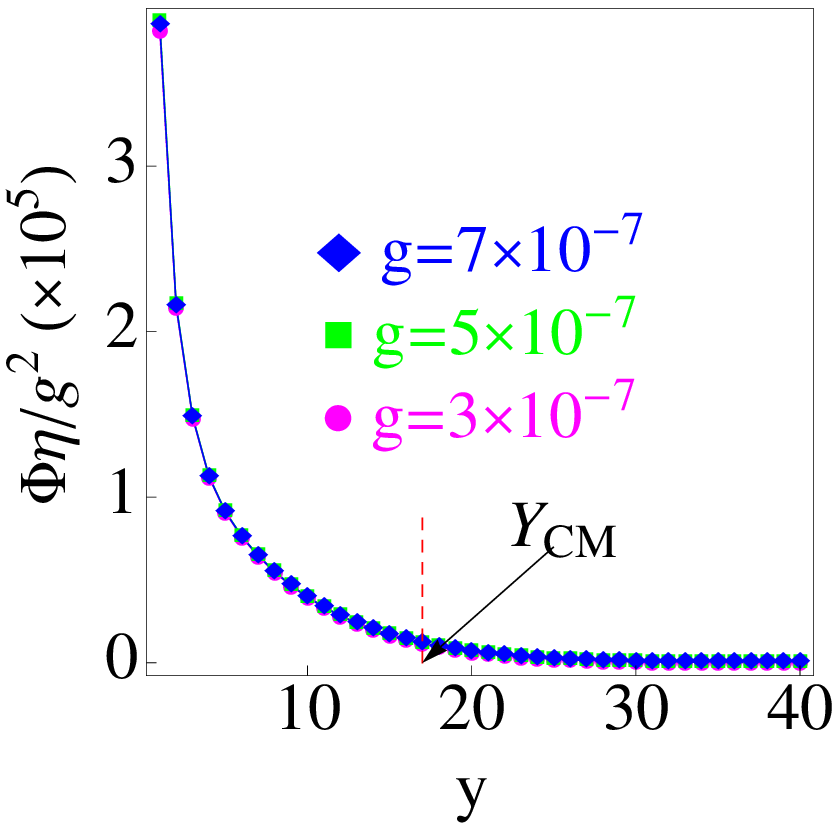}
\caption{Left: Dissipation rate integrated along a horizontal line at $y$, $\Phi(y)$. Right: The same quantity as in the left panel but divided by $g^{2}$. The observed master curve supports the validity of \equref{eq:PhiT}.}
\label{fig_dissipation_collapsed_forcing}
\end{figure}

\begin{figure}[!ht]
\includegraphics[ width=6cm]{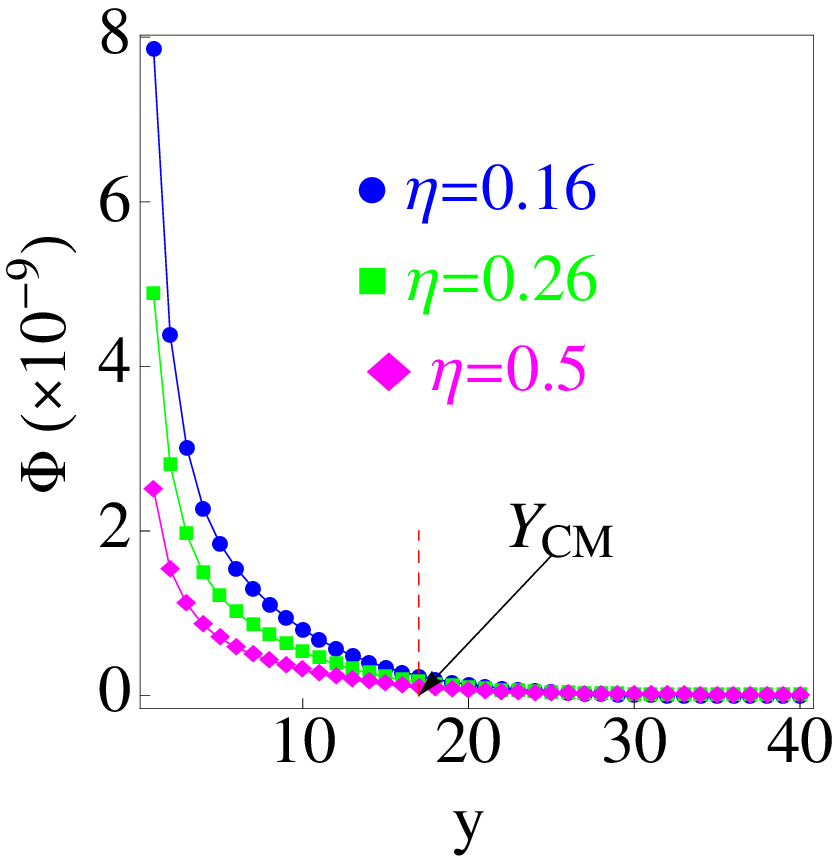}\hspace*{5mm}
\includegraphics[ width=6cm]{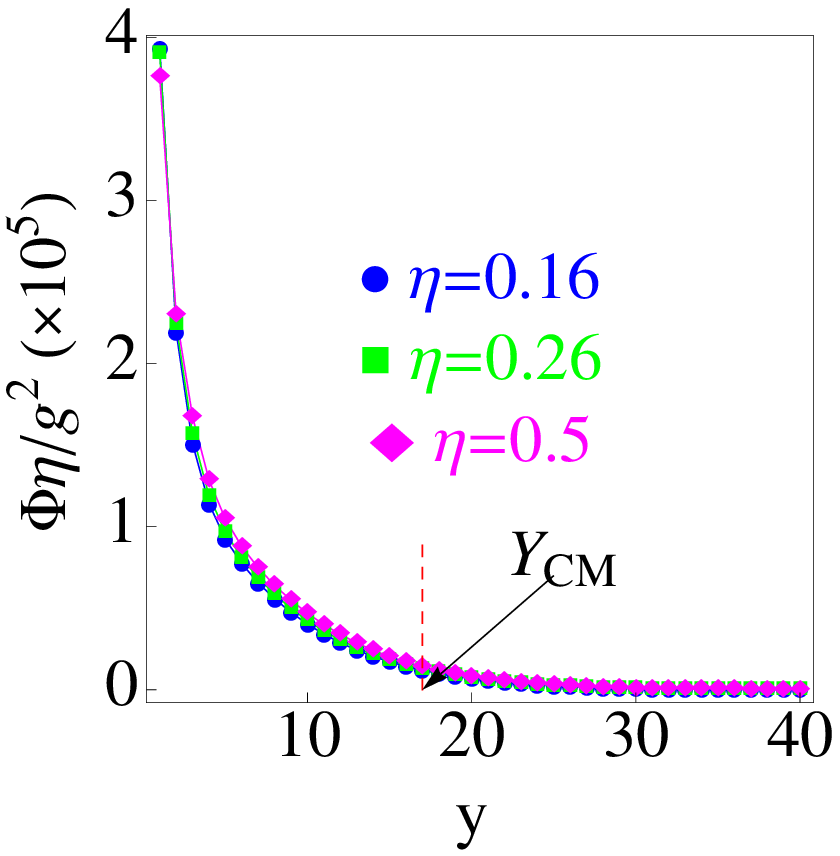},
\caption{A similar plot as in \figref{fig_dissipation_collapsed_forcing} but now for various fluid viscosities $\eta$. Here, the right panel depicts $\Phi(y)$-data from the left panel multiplied by $\eta$. Again, the validity of \equref{eq:PhiT} is supported by the master curve.}
\label{fig_dissipation_collapsed_viscosity}
\end{figure}

In order to verify \equref{eq:PhiT}, we determine $\Phi(y)$ for different values of body force and viscosity. Typical plots of the thus obtained results are shown in Figs.~\ref{fig_dissipation_collapsed_forcing} and \ref{fig_dissipation_collapsed_viscosity}. These data clearly underline the validity of \equref{eq:PhiT} within the studied range of parameters. 

In addition to supporting the validity of \equref{eq:PhiT}, the data shown in Figs.~\ref{fig_dissipation_collapsed_forcing} and \ref{fig_dissipation_collapsed_viscosity} provide further evidence for the fact that most part of dissipation occurs in the region below the droplet's center-of-mass. Based on this observation, we propose a simple relation allowing to describe the dependence of droplet velocity on contact angle.

Our simple analytic model is based on scaling arguments. To proceed, we start with the energy balance equation for a cylindrical droplet of axial length $L_x$ in the steady state. Using the translation invariance with respect to the $x$-coordinate, one can write $g \rho \pi \Reff^2 L_x \Ucm = L_x (\eta/2) \int_0^H \int \dot{\gamma}^2 dz dy$, where $\dot{\gamma}$ is the local shear rate. Since the energy is almost completely  dissipated in a region below $\Ycm$, we can safely restrict the upper limit of the integration to $\Ycm$  and rewrite the energy ballance equation as $2 g \rho \pi \Reff^2 \Ucm=\eta \int_{0}^{\Ycm} \int \dot{\gamma}^2 dzdy$ (see \figref{fig_viscous_dissipation_y}). Neglecting droplet deformation, the droplet's cross-section is a circular segment with a base contact angle of $\thetaeq$. Here, we assume that $\dot{\gamma}$ simply scales as $\Ucm/\Ycm$ throughout the droplet. This may appear as a crude approximation, but it allows to obtain a solvable analytic expression.  Furthermore, we approximate the surface of the droplet below $\Ycm$ as that of a rectangle of height $\Ycm$ and length $l_z$. Adopting this, the right hand side of the energy balance equation can now be estimated by  $(\eta/2) l_z \Ycm \times (\Ucm/\Ycm)^2=(\eta/2) l_z \Ucm^2/\Ycm$. Thus, one obtains $2g \rho \pi \Reff^2 \Ucm = \eta l_z \Ucm^2/\Ycm$, which then yields $\Ucm=g \rho \pi \Reff^2 \Ycm/(l_z \eta)$. For the considered geometry, the quantity $\Ycm/l_z$,  is only a function of $\thetaeq$. Taking this into account, we finally arrive at
\begin{equation}
\Ucm =C\frac{g \rho\Reff^2}{\eta}\left[\dfrac{4\text{sin}^{2}\thetaeq}{3(2\thetaeq-\text{sin}2\thetaeq)}-\text{cot}\thetaeq\right].
\label{eq:Ucm_vs_thetaY}
\end{equation}

\begin{figure}[!ht]
 \includegraphics[ width=8cm]{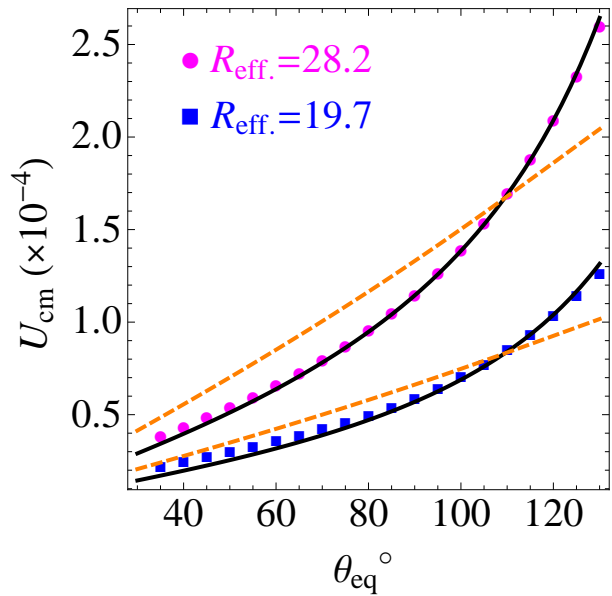}
\caption{Droplet velocity versus equilibrium contact angle for two different droplet volumes as indicated. Full solid lines are best fit results to  \equref{eq:Ucm_vs_thetaY} while dashed lines give best fit results to  $ \Ucm =C (g \rho\Reff^2/\eta)(1-\text{cos}\thetaeq)^{2}/(\thetaeq-\text{sin}\thetaeq\text{cos}\thetaeq)$ (see the text).  The force density and dynamic viscosity are fixed to $g=10^{-7}$ and  $\eta=0.16$ respectively  for both droplets.}
 \label{velocity_theta_eq}
\end{figure}

The validity of the model has been tested in \figref{velocity_theta_eq} for two different droplet radii.  This simple model reproduces well the simulation results. Interestingly, the fitting prefactor, $C$, for both investigated droplet sizes are very close to each other ($0.56$ and $0.57$) showing the consistency of the model.

We would like to emphasize that the present approach is different from conventional approaches, where the integration is taken over the entire volume of droplet. Following the conventional route, one would obtain  a different expression,  $ \Ucm =C (g \rho\Reff^2/\eta)(1-\text{cos}\thetaeq)^{2}/(\thetaeq-\text{sin}\thetaeq\text{cos}\thetaeq)$,  which, as  shown in \figref{velocity_theta_eq}, is  not successful in capturing the observed behavior.

It is noteworthy  that an extension of \equref{eq:Ucm_vs_thetaY} to $3D$ can simply be obtained by writing the energy ballance equation in $3D$ and replacing the corresponding expression for $\Ycm/l_z$ by that of a spherical cap.

\section{Conclusion}
We use a two-phase lattice Boltzmann method to study the dynamics of cylindrical droplets on a flat substrate under the action a gravity-like external force density. Starting from the energy ballance equation, we first drive a simple analytic relation, \equref{eq:Ucm}, indicating that ---as long as the shape-invariance of droplet is maintained--- droplet's center-of-mass velocity, linearly scales with force density, and the square  of the droplet radius.  At strong body forces or large droplet volumes, deviations from \equref{eq:Ucm} are observed. A survey of droplet shape within our simulations suggest that droplet deformation is indeed the main cause of observed deviations from the simple scaling relation. Interestingly, however, the droplet's center-of-mass velocity remains proportional to the inverse of the dynamic viscosity regardless of droplet's deformation state. This is in line with the idea that viscosity merely affects the time scale of the problem with no influence on droplet shape. A detailed study of the local dissipation inside droplet is also provided. A results of these investigations is that dissipation mainly occurs close to the three phase contact line and within a region below the droplet's center-of-mass. Using the latter observation, we propose a simple analytic expression accounting for the dependence of droplet velocity on the equilibrium contact angle. Results of computer simulations confirm the validity of this simple model.\\

\section{Acknowledgments}
We would like to thank Dmitry Medvedev and Markus Gross for insightful discussions. M.G. is also acknowledged for providing us a version of his LB code. N.M. gratefully acknowledges the grant provided by the Deutsche Forschungsgemeinschaft (DFG) under the number Va 205/3-2. ICAMS gratefully acknowledges funding from ThyssenKrupp AG, Bayer MaterialScience AG, Salzgitter Mannesmann Forschung GmbH, Robert Bosch GmbH, Benteler Stahl/Rohr GmbH, Bayer Technology Services GmbH and the state of North-Rhine Westphalia as well as the European Commission in the framework of the European Regional Development Fund (ERDF).

%%\bibliographystyle{prsty_with_title}
%%\bibliography{literature_2011_01_11}

%----------------------------------------------------------------

\end{document}